\documentclass[a4paper, superscriptaddress, sd, amsfonts, amssymb, amsmath, preprint, showkeys, nofootinbib, nobibnotes]{revtex4-1}
\usepackage[english]{babel}
\usepackage[utf8]{inputenc}

\usepackage{xcolor}
\usepackage{graphicx}
\graphicspath{{../RSC-PhiloTransA/},{./}} 
\usepackage{dcolumn}
\usepackage{bm}
\usepackage{nicefrac}
\usepackage{units}
\usepackage{textcomp}
\usepackage[]{natbib} 
\usepackage{hyperref}
\hypersetup{colorlinks=true, citecolor=blue} 
\usepackage{cleveref}
\usepackage{verbatim}
\crefname{section}{§}{§§}
\Crefname{section}{§}{§§}
\usepackage{soul}
\usepackage[normalem]{ulem}

\begin{document}
\title{Lattice Boltzmann simulations of drying suspensions of soft particles}

\author{M. Wouters} 
\affiliation{Department of Applied Physics, Eindhoven University of Technology, De Rondom 70, 5612 AP, Eindhoven, The Netherlands}
\affiliation{Helmholtz Institute Erlangen-N\"{u}rnberg for Renewable Energy, Forschungszentrum J\"{u}lich, F\"{u}rther Stra{\ss}e 248, 90429 N\"{u}rnberg, Germany}
\author{O. Aouane} 
\affiliation{Helmholtz Institute Erlangen-N\"{u}rnberg for Renewable Energy, Forschungszentrum J\"{u}lich, F\"{u}rther Stra{\ss}e 248, 90429 N\"{u}rnberg, Germany}
\author{M. Sega} 
\affiliation{Helmholtz Institute Erlangen-N\"{u}rnberg for Renewable Energy, Forschungszentrum J\"{u}lich, F\"{u}rther Stra{\ss}e 248, 90429 N\"{u}rnberg, Germany}
\author{J. Harting}
\email{j.harting@fz-juelich.de}
\affiliation{Helmholtz Institute Erlangen-N\"{u}rnberg for Renewable Energy, Forschungszentrum J\"{u}lich, F\"{u}rther Stra{\ss}e 248, 90429 N\"{u}rnberg, Germany}
\affiliation{Department of Chemical and Biological Engineering and Department of Physics, Friedrich-Alexander-Universit\"at Erlangen-N\"urnberg, F\"{u}rther Stra{\ss}e 248, 90429 N\"{u}rnberg, Germany}

\date{\today} 

\begin{abstract}
The ordering of particles in the drying process of a colloidal suspension is crucial in determining the properties of the resulting film. For example, microscopic inhomogeneities can lead to the formation of cracks and defects that can deteriorate the quality of the film considerably. This type of problem is inherently multiscale and here we study it numerically,  using our recently developed method for the simulation of soft polymeric capsules in multicomponent fluids. 
We focus on the effect of the particle softness on the film microstructure during the drying phase and how it relates to the formation of defects. We quantify the order of the particles by measuring both the Voronoi entropy and the isotropic order parameter. Surprisingly, both observables exhibit a non-monotonic behaviour when the softness of the particles is increased. We further investigate the correlation between the interparticle interaction and the change in the microstructure during the evaporation phase. We observe that the rigid particles form chain-like structures that tend to scatter into small clusters when the particle softness is increased.
\end{abstract}

\maketitle

\section{Introduction}\label{sec:introduction}
The dynamics of drying suspensions have sparked the scientific community's interest for the last two decades due to the wide range of phenomena observed during the drying and their end-user applications in areas spanning from household products to manufacturing processes~\cite{Keddie2010,Kooij2015}. In addition to their aesthetic function, paints and coatings' primary function is to provide the treated substrate with protection from external factors such as weather conditions, mechanical damage, and corrosion. The viscosity and the microstructure of the suspension during the solvent evaporation phase, when stresses build up in the film, are crucial in determining the paint's properties and can lead to cracks and defects in the final dry paint~\cite{Dufresne2003,Singh2007,Roberts2012,Goehring2013,Man2008,Piroird2016}. The occurrence of these unwanted phenomena, however, can be mitigated by exploiting the sensitivity of the suspension's microstructure which is sensitive to the softness and deformability of the suspended particles~\cite{Winnik1997,Routh2001,Keddie2010}.

Drying is a complex process, and computational approaches that can model the deformation and interactions of many soft particles during the drying of a suspension can be of great value to better understand the different aspects of the drying. Given the spatial and temporal scales involved, microscopic methods can only give partial insight into the problem~\cite{MHS16}. Theoretical models such as the diffusive model~\cite{Trueman2012}, on the other hand, give insight in average properties and particle densities, but cannot incorporate the complex interactions due to the softness of the densely packed particles. Several mesoscopic methods have been developed to circumvent these limitations~\cite{Fujita2007,Fujita2015,Uzi2015,Zhao2019,Liang2013}.
However, to the best of our knowledge, no computational models are present, which can model the evaporation of a suspension at the mesoscale and efficiently simulate the influence of the softness on the drying process suspension of thousands of particles.

Here, we focus on the drying of a suspension consisting of soft colloidal particles and apply our recently introduced simulation method~\cite{Wouters19,WASH20} to study the influence of the particle softness and deformability on the ordering and structuring during drying. 
In this sense, we remain in the domain of generic, drying colloidal suspensions of soft particles, without treating specific aspects of latex paints like the polymeric nature of the particles and the latex film-formation. For a drying latex paint, this corresponds to simulating the drying process up to the point where the coalescence of the particles does not yet play a significant role.

\section{Numerical Method}\label{sec:method}
The Navier-Stokes equations can be recovered in the incompressible limit using the discretized Boltzmann equation (LBM) in velocity space where we track the time evolution of the particle distribution function $f_i(\mathbf{x},t)$ on a simple cubic lattice corresponding to the probability of finding a fluid particle moving with a velocity $\mathbf{e}_i$, located at a position $\mathbf{x}$, and at time $t$ such that
\begin{equation}
   f_i^c(\mathbf{x} + \mathbf{e}_i \Delta t , t + \Delta t)-f_i^c(\mathbf{x},t) = \frac{-\Delta t} {\tau^c} \bigg[  f_i^c(\mathbf{x},t) -f_i^\mathrm{eq}(\mathbf{x},t)\bigg]
    \mbox{,}
    \label{eq:LBM}
\end{equation}
where the subscript $i=1,19$ refers to the set of possible discrete velocities $\mathbf{e}_i$ on the lattice, while the superscript $c$ is used to identify between the different fluid components~\cite{benzi1992lattice,kruger2017lattice}. $\tau$ is the relaxation time related to the kinematic viscosity such that $\nu^c = c_s^2[\tau^c - \frac{\Delta t} {2}]$ with $c_s=\frac{1}{\sqrt{3}}\frac{\Delta x}{\Delta t}$ being the lattice speed of sound, $\Delta t$ the discrete time step, and $\Delta x$ the lattice spacing. The RHS in equation \ref{eq:LBM} corresponds to the standard Bhatnagar-Gross-Krook (BGK) collision operator \cite{bhathnagor1954model}. $f_i^{\mathrm{eq}}$ is the local equilibrium distribution function corresponding to the truncated expansion of the Maxwell-Boltzmann distribution for the velocities in an ideal gas which reads as
\begin{equation}
  f_i^{\mathrm{eq}} = \omega_i \rho^c \bigg[ 1 + \frac{\mathbf{e}_i \cdot \mathbf{u}^c}{c_s^2} - \frac{ \left( \mathbf{u}^c \cdot \mathbf{u}^c \right) }{2 c_s^2} + 
  \frac{ \left( \mathbf{e}_i \cdot \mathbf{u}^c \right)^2}{2 c_s^4}  \bigg]
  \mbox{.}
  \label{eq:f_equilibrium}  
\end{equation}
Here, $\omega_i$ is a set of weights defined as $1/3$, $1/18$ and $1/36$ for $i=1$, $i=2\dots7$, and $i=8\dots19$, respectively for the D3Q19 lattice used in this work~\cite{qian1992lattice}.
Forces acting on the fluid are included as a shift in the
equilibrium velocity in Eq.~\ref{eq:f_equilibrium}~\cite{shan1993lattice}.
The local macroscopic density $\rho^c$ and velocity
$\mathbf{u}^c$ can be computed as
\begin{align}
    \rho^c(\mathbf{x}) &= \rho_0 \sum_i f_i^c(\mathbf{x}), \\
\rho^c(\mathbf{x})\mathbf{u}^c(\mathbf{x}) &= \rho_0 \sum_i f_i^c(\mathbf{x}) \mathbf{c}_i + \frac{\rho_0\Delta t}{2} {\mathbf{F}^c(\mathbf{x})}.
\end{align}
For simplicity, we keep the reference density $\rho_0$, $\Delta t$, $\Delta x$ and $\tau$ at unity for the remainder of this paper. 

To account for the interaction between the different components, we use the  pseudo-potential interaction scheme of Shan and Chen~\cite{shan1993lattice}, where the interaction force $\mathbf{F}^c$ acting between fluid components on neighbouring fluid nodes along $c_i$ reads as
\begin{equation}
    \mathbf{F}^{c}(\mathbf{x}, t) = -\psi^c(\mathbf{x}, t) \sum\limits_{c'}G^{cc'}\sum\limits_i \omega_i \psi^{c'}(\mathbf{x}+\mathbf{e}_i, t) \mathbf{e}_i.
    \label{eq:shanchen}
\end{equation}
Here, $G^{cc'}$ is the interaction parameter, $\psi^c$ is the pseudo density of the pseudo-potential defined as $\psi^c(\mathbf{x}, t) = \rho_0 [1-e^{-\rho^c(\mathbf{x},t)/\rho_0}]$.
The fluid interaction force $\mathbf{F}^c$ is included in equation \ref{eq:LBM} as a shift in the equilibrium velocity of component $c$ as discussed in \cite{shan1993lattice,Wouters19}.

The evaporation is emulated by modifying the densities at the evaporative boundary while enforcing the mass conservation of the total system by locally adding the removed mass of component $c$ at the evaporative boundary to component $c^\prime$ in accordance with Fick's law.
For each fluid node $\mathbf{x}_{ev}$ on the evaporative boundary layer, the particle distribution function is modified such that
\begin{equation}
    f_i^c(\mathbf{x}_\mathrm{evap},t) = f_i^{eq}(\rho_\mathrm{evap}^{c},\mathbf{u}^{c}(\mathbf{x}_\mathrm{evap},t)),
    \label{eq:evap}
\end{equation}
with $\mathbf{u}^{c}(\mathbf{x}_\mathrm{evap},t)$ and $\rho_\mathrm{evap}^{c}$ being parameters controlling the velocity and the density at the evaporative boundary, respectively. The choice of $\rho_\mathrm{evap}^{c}$ with respect to the minority density can lead to either the condensation or the evaporation of the $c$-th component. A schematic of the model is shown in figure S1.
The evaporation model has been validated in our previous paper \cite{hessling2017diffusion} by measuring the evolution of the interface height ($h$) during the evaporation. The numerical and analytical results showed a good quantitative agreement.


Our particles are modelled as elastic closed membranes enclosing an inner fluid and known as capsules. The membrane is endowed with resistance to shear elasticity, area dilatation, and bending.
Under the assumption of membranes with negligible thickness compared to their radius, the total strain energy of the membrane can be modelled with a 2D hyperelastic constitutive law such as the Skalak law \cite{skalak1973strain} given by
\begin{equation}
    E_s = \int_A \frac{\kappa_s}{4}[I_1^2 + 2 I_1 -2 I_2 + C I_2^2] dA, 
    \label{eq:strain_energy}
\end{equation}
where $\kappa_s$ is the shear elastic modulus, $C = \frac{1}{2}[\frac{\kappa_a}{\kappa_s}-1]$ is a dimensionless parameter describing the strain hardening nature of the membrane, and $\kappa_a$ is the area dilatation modulus. The surface of the membrane is discretized with a triangular mesh, where $N_f = 1280$ and $N_n = 720$ are the number of faces and nodes, respectively, used throughout this work.  
No-slip boundary conditions are applied on the surface of the membrane, {{and the capsule is assumed to be incompressible, so to say, its volume has to be conserved over time. The incompressibility condition is enforced by using a penalty function that controls the deviation of the capsule volume ($V$) during the simulation from its prescribed reference volume ($V_0$). The expression of the penalty function ($E_v$) is given by 
\begin{equation}
    E_v = \frac{\kappa_v}{2} \frac{(V-V_0)^2}{V_0},
    \label{eq:volume_conservation}
\end{equation}
where $\kappa_v$ is a free parameter controlling the volume conservation of the capsule.}}
The force on each membrane node $\mathbf{x}_{i}^{m}$, with the superscript $m$ referring here to the membrane, is obtained using the principle of virtual work such that
\begin{equation}
    \mathbf{F}_\alpha^m(\mathbf{x}_{i}^{m}) = -\frac{\partial E_\alpha}{\partial \mathbf{x}_{i}^{m}},
\end{equation}
where the subscript $\alpha$ denotes either $s$ for the strain force or $v$ for the force corresponding to the penalty function on the volume. The elastic force is evaluated using a linear finite element method (FEM) as discussed in \cite{kruger2011efficient}.
The membrane bending force stems from the functional derivative of the Helfrich free energy that reads as 
\begin{equation}
    \mathbf{F}_b^m(\mathbf{x}_{i}^{m}) = 2 \kappa_b [2H \{H^2 - K\} + \Delta_s H]\mathbf{n},
\end{equation}
where $k_b$ is the bending modulus, $H = \frac{(\hat{c}_1+\hat{c}_2)}{2}$ is the mean curvature with $\hat{c}_1$ and $\hat{c}_2$ being the two principal curvatures, $K = \hat{c}_1\hat{c}_2$ is the Gaussian curvature, $\Delta_s$ is the surface gradient, and $\mathbf{n}$ is the unit outward normal vector. The discretization of the surface gradient, and the main and Gaussian curvatures follows the approach based on the discrete differential geometry operators as introduced in \cite{meyer2003discrete,biben2011three}.

The equilibrium shape of the particle is governed by two non-dimensional numbers. The first one is the two-dimensional Poisson ratio defined for Skalak capsules as $\nu_{s} = C/(1+C)$ with $\nu_s \in ]-1,1]$ and $C$ being a dimensionless parameter defined as the ratio between the area dilatation and the shear elasticity of the membrane \cite{barthes2016motion}.
The second dimensionless number is the softness parameter ($\beta$) which quantifies the softness of the particle with respect to the surface tension $\gamma$ of the fluid-fluid interface \cite{Wouters19}. $\beta$ is expressed as
\begin{equation}
    \beta = \frac{R_0^2\gamma}{\kappa_b},
\end{equation}
where $R_0$ is the radius of the undeformed particle. 
\begin{figure}[h!]
    \centering\includegraphics[width = 1.0\textwidth]{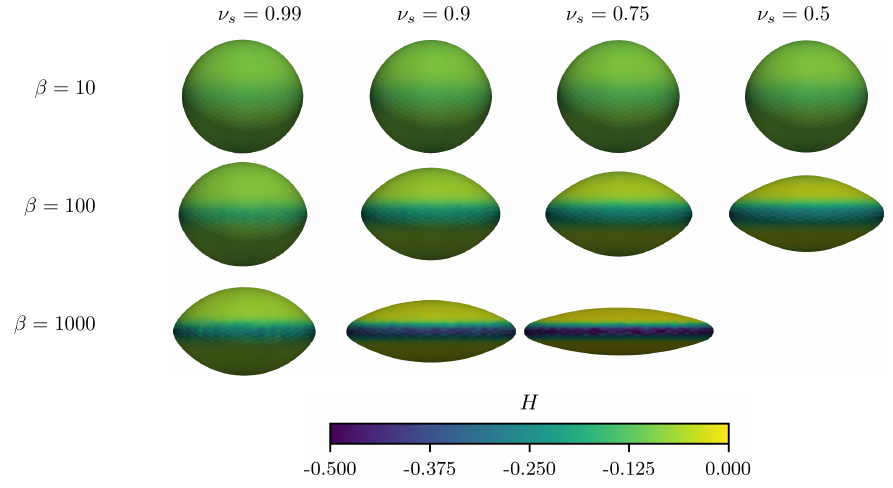}
    \caption[]{Equilibrium shapes for an initially spherical particle adsorbed onto a fluid-fluid interface for various softness parameters ($\beta$) and Poisson ratios ($\nu_{s}$). The colours depicts the local mean curvature ($H$).
        \label{fig:single_capsule}
        }
\end{figure}
Figure \ref{fig:single_capsule} shows the effect of $\nu_{s}$ and $\beta$ on the equilibrium shape of a capsule adsorbed onto a fluid-fluid interface. We observed that at the interface the capsule behaves qualitatively as a drop for $\beta \gg 1$ and as a quasi-rigid particle for $\beta \ll 1$.
Throughout this paper, $\nu_{s}$ is fixed to $0.9$. The role of $\nu_{s}$ is discussed in the supplementary material.

To ensure a sufficiently large gap size between two particle boundaries, we use a short-range Hertz interaction potential. When the distance $\delta$ between a boundary node and another particle or the substrate is below the interaction cut-off distance $\delta_0$, we apply a force of respectively

\begin{align}
    & \mathbf{F}^{\text{PP}}(\delta)    = \Xi^{\text{PP}} \delta^{1.5} \hspace{.5cm} \text{if} \hspace{.25cm} \delta < \delta_0^{\text{PP}},\\
    & \mathbf{F}^{\text{PS}}(\delta)    = \Xi^{\text{PS}} \delta^{1.5} \hspace{.5cm} \text{if} \hspace{.25cm} \delta < \delta_0^{\text{PS}}.
    \label{eq:fluidstructure:interactions}
\end{align}

The fluid-interface coupling method used in this paper is based on the exchange of momentum between the fluid and the membrane by applying the half-way bounce-back rules (SBB) for distribution functions $f^c_i$ that would cross a boundary element during the streaming step. We follow here the exact same procedure we introduced in detail in our recent method paper here we studied the deformation of a soft particle adsorbed at a fluid-fluid interface~\cite{Wouters19}.

The particles are advected in time following a forward Euler scheme where each membrane node position is updated such that 
\begin{equation}
    \mathbf{x}^{m}_i(t+\Delta t) = \mathbf{x}^{m}_i(t) + \Delta t\frac{\mathbf{F}_{{tot},i}^m}{M_i}.
\end{equation}
Here, $M_i$ is the mass of the $i$-th membrane node, and $\mathbf{F}_{{tot},i}^m$ is the total nodal force which includes the contributions of the elastic, bending, particle-particle and particle-substrate forces.

For more details on our implementation we refer the reader to some of our previous publications~\cite{Wouters19,WASH20,wouters2020mesoscopic,hessling2017diffusion,KKH14,HKH11,HVC04,SSFKSHC17,ASH20}.

\section{Particle ordering in a single layer of soft particles}\label{sec:relaxation}

During the drying process, the particles become tightly packed at the liquid interface. Defects in the hexagonal close-packing structure can occur due to many external factors, and are extremely important in determining the quality of the dried film, as these inhomogeneities can lead to the development of cracks.  In this section, we study the relaxation of particles in a thin fluid film on a substrate and the characteristics of the defects that are formed during this process, as a function of the particle softness. We initialise a domain of size $D=1000\times{}1000\times{}30$ lattice units, with a fluid film of one component (``red'') and a thickness of $h=R_0+\delta_0$ (i.e., such that no significant deformation of the fluid interface occurs during the simulation) on top of a rigid substrate. The rest of the domain is filled with a second fluid component (``blue''). We set the particle fraction to a value $\Phi$ close to the close-packing fraction $ \Phi_0 = \pi \sqrt{3}/6$ of circles, using a radius of $\hat{R} + 0.5\delta_0$. Here, $\hat{R}$ represents the equilibrium radius of a single particle adsorbed onto the fluid-fluid interface $\hat{R}$, so that with our choice all particles can be placed at the fluid-fluid interface. 

To determine the degree of local order, we use the Voronoi tessellation of the (instantaneous) positions of the centres of mass of the particles. 
\begin{figure}
    \centering\includegraphics[width = .75\textwidth]{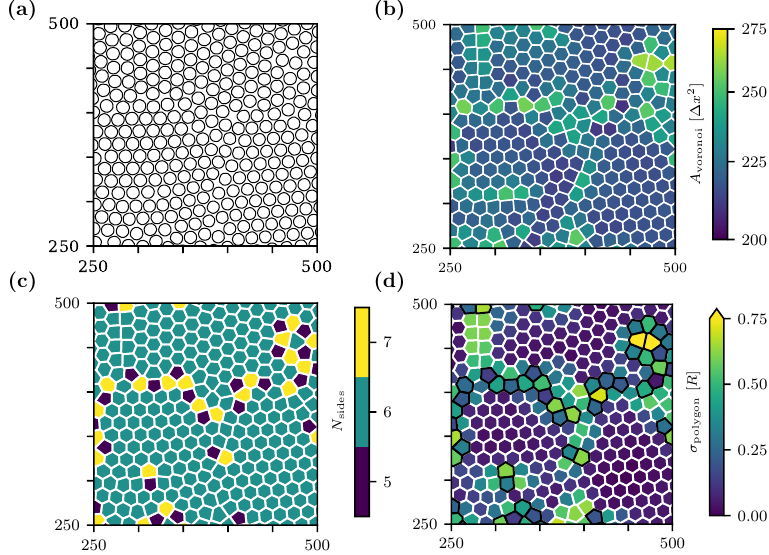}
    \caption[Voronoi tessellation of thin film of soft particles]{ 
    Comparison between the particle boundaries and the Voronoi tessellation for the system with 4445 particles and $\beta=25$ at $t=10^5$.
    Only 1/16 of the simulated domain is visualised for the sake of clarity. 
    (a) Particle boundaries; 
    (b) Voronoi tessellation, with polygons coloured based on their area;
    (c) Voronoi tessellation, with polygons coloured based on their number of sides.
    (d) Voronoi tessellation, with polygons coloured based on the standard deviation of the side length of the polygon $\sigma_{\mathrm{polygon}}$.
    Polygons with a number of sides other than 6 are outlined in black.
        \label{fig:voronoi-example}
        }
\end{figure}
In figure~\ref{fig:voronoi-example}a we show a typical snapshot of a film containing 4445 particles with $\beta=25$ after $10^5 \Delta t$. The particle boundaries are computed in a small horizontal slice taken at the mean height of the centre of mass of all particles.
In some regions, the particles are evenly packed in hexagonal structures, whereas different types of defects can be observed in the remaining regions.
For example, in the top left corner, we observe a vertical line defect, along which the particles are close to a square packing, instead of the hexagonal one. Near the centre, and at the centre in the bottom we observe several defects characterised by a considerably large area per particle. Also, just below the top right corner, a ring-like configuration of 5 particles is present, with a significant gap between the particles.

Figure~\ref{fig:voronoi-example}b shows the corresponding Voronoi tessellation, where the polygons are coloured based on the Voronoi area.
The Voronoi diagram shows quite regular hexagons in a large part of the simulation cell and distorted polygons in the regions where the particle packing shows defects or non-hexagonal packing. For example, the line defect in the top left corner results in polygons which are close to squares on the side of the line defect while being hexagonally shaped on the opposite side. This change in shape results in an increase of the corresponding area of the polygons.

Around a horizontal line defect at roughly two-thirds of the visualised domain, the particles show a varying order with sometimes significantly larger inter-particle gaps, which translates into an increase in the Voronoi areas of the surrounding polygons as seen in Fig.~\ref{fig:voronoi-example}b, and a deviation from a six-sided polygon.

A typical quantifier for how well a system is ordered in a homogeneous packing is the Voronoi entropy, defined as~\cite{Limaye1996}
\begin{equation}
    S_{\mathrm{vor}} = - \sum\limits_n P_n \mathrm{ln}(P_n), \ \ n \ge 3,
\end{equation}
where $P_n$ is the fraction of polygons with $n$ sides or edges.
For a random 2D distribution, one would expect a value $S_{\mathrm{vor}}=1.71$, which decreases as the particle packing becomes more homogeneously ordered. However, the deviation for our expected equilibrium six-sided polygon is not always present near defects, where for instance the polygons near the vertical line defect in the top left corner are still six sided, where one of the polygon sides is very small compared to the other sides.
This is clearly seen in figure~\ref{fig:voronoi-example}c, where the polygons are coloured based on the number of sides $N_{\mathrm{sides}}$.

In figure~\ref{fig:voronoi-example}d the polygons are coloured based on the standard deviation of the side lengths $\sigma_{\mathrm{polygon}}$.
Here, we indeed see a significant increase in the standard deviation in the regions where the particles are not hexagonally packed, or where the typical spacing between the particles is not homogeneous. In addition to the Voronoi entropy, we monitor also the isotropic order parameter, defined as~\cite{Liang2013}
\begin{equation}
    IOP = \frac{n_{\mathrm{equi}}}{N},
\end{equation}
where $n_{\mathrm{equi}}$ is the number of equilateral polygons of the Voronoi tessellation, and $N$ is the total number of Voronoi polygons (i.e., the total number of particles).
Here, we consider a polygon to be equilateral when the standard deviation of the length of its sides  ${\sigma_{\mathrm{polygon}}} \leq 0.08{\hat{R}}$~\cite{Liang2013}, where the effective radius ${\hat{R}}={R_0}+0.5{\delta_0}$ corrects the initial radius for the effect of the particle-particle interaction range.
The ${IOP}$ is close to $0$ for disordered structures, and approaches $1$ for highly ordered structures.

After the simulation is started, the system relaxes and the soft particles are driven towards the hexagonal ordering.
Since the particles are adsorbed at the interface of the thin fluid film, there is only a minimal movement in the vertical direction perpendicular to the substrate, which we neglect for now.
During the relaxation regions are formed where most particles are packed according to a hexagonal packing, while different types of defects persist between these regions, as was already seen above.

\begin{figure}
    \centering\includegraphics[width = .75\textwidth]{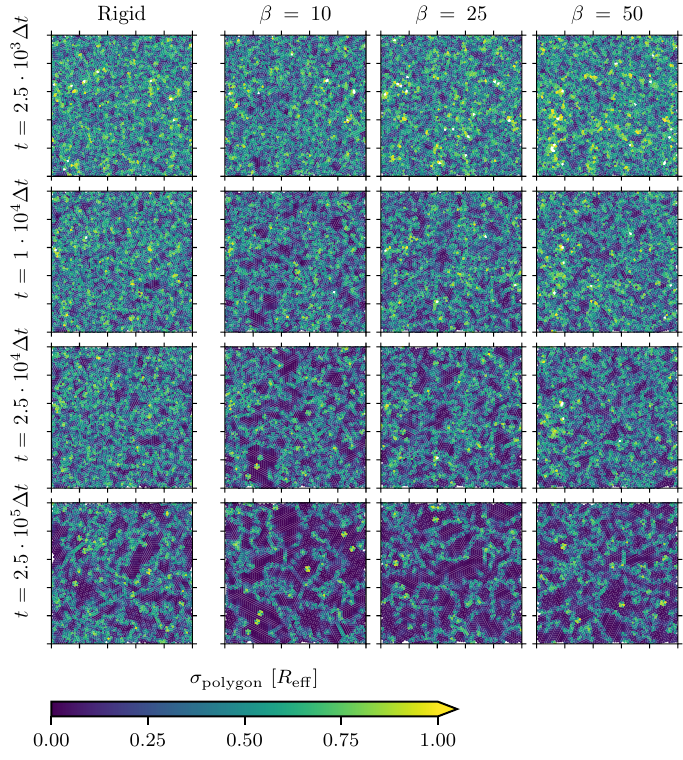}
    \caption[Qualitative visualisation of isotropic ordering in a layer of soft particles]{
        Top view of the Voronoi tessellation of a single layer of soft particles
        adsorbed onto the thin fluid interface supported by the substrate,
        for the same system with 4445 particles as in Fig.\ref{fig:voronoi-example}.
        The polygons are coloured according to $\sigma_\textrm{polygon}$, with darker regions characterizing ordered areas and brighter regions the disordered neighborhood of defects.
        Each tick mark indicates a distance of $200\Delta x$. The small circular regions with a high $\sigma_\textrm{polygon}$  occur when particles are trapped in a ring-like configuration without a particle in the centre of the ring.
    }
    \label{fig:voronoi-equilateral}
\end{figure}

In figure~\ref{fig:voronoi-equilateral} we visualise the Voronoi tessellation coloured by ${\sigma_{\mathrm{polygon}}}$ (c.f. figure~\ref{fig:voronoi-example}d).
It can already be qualitatively observed that stiffer particles form regions with a hexagonal packing faster than softer particles.
However, perfectly rigid particles reach a hexagonal packing much slower than the soft particles with ${\beta}=50$.

This behaviour can be understood by considering that the elasticity of the capsules provides additional degrees of freedom, increasing the number of pathways for the energy to relax and reach its minimum. On the contrary, rigid particles, having no internal degree of freedom can be brought to the energy minimum only by rearranging their relative positions. We have attempted in our previous work \cite{WASH20} to correlate the deformation energy of the particles to the fluid one by measuring the capillary force between two particles adsorbed onto a fluid-fluid interface for different softness parameters ($\beta$) and comparing the results to the rigid particle case. We observed that the capillary interaction force decreases as the softness parameter increases.

\begin{figure}
    \centering\includegraphics[width = .9\textwidth]{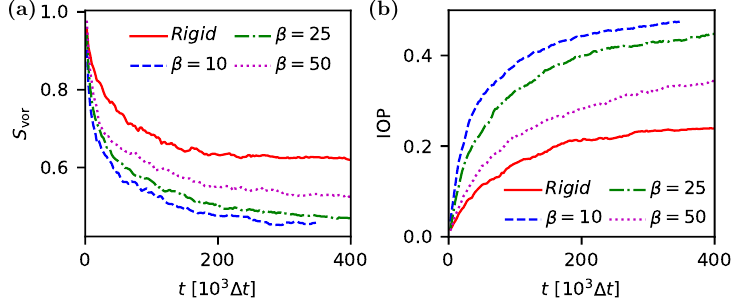}
    \vspace{-3mm}
    \caption[Order parameters for a layer of soft particles]{
        Relaxation of the Voronoi entropy (left panel) and of the isotropic order parameter (right panel) of a single layer of densely packed soft particles for the same system with 4445 particles as in Fig.\ref{fig:voronoi-example}.
    }
    \label{fig:order-parameter}
\end{figure}

In figure~\ref{fig:order-parameter}a we show the evolution of the Voronoi entropy for systems with different softness parameters.
Here, it can be observed that for softer particles the decay of the Voronoi entropy is slower than for stiffer particles, whereas for perfectly rigid particles it decays much slower and appears to settle at a value around 0.6.
Similarly, the isotropic order parameter as shown in Fig.~\ref{fig:order-parameter}b increases faster for stiffer particles than for softer particles, while perfectly rigid particles again order at a greatly diminished rate as compared to the relatively stiff particles with ${\beta}=10$.

\begin{figure}
    \centering\includegraphics[width = .75\textwidth]{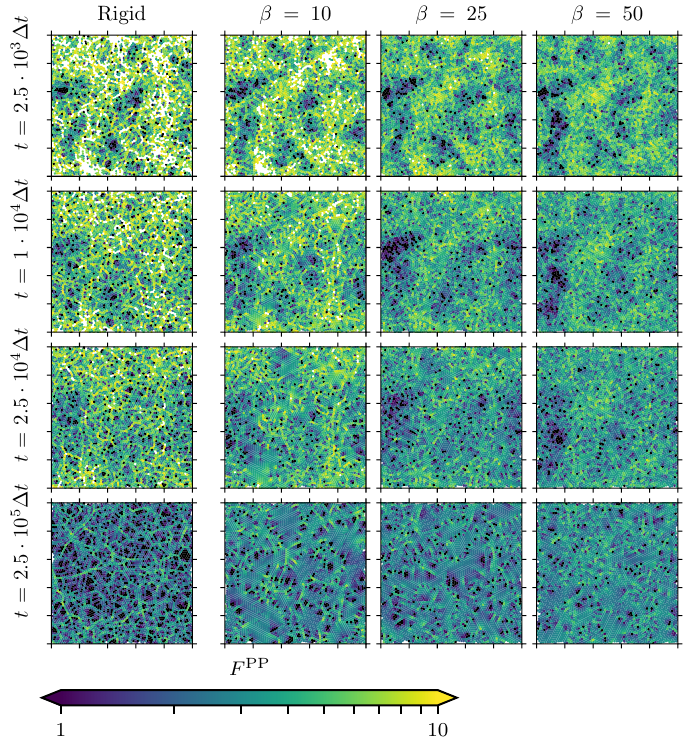}
    \caption[Interaction forces in a layer of soft particles]{
        Top view of the Voronoi tessellation, with polygons coloured according to the magnitude of the total  force,
        for the same system with 4445 particles as in Fig.\ref{fig:voronoi-example}.
    After ${1.5\times{}10^5}$ simulation steps the rigid particles have relatively more hexagonal-like structures, as compared to softer particles.}
    \label{fig:voronoi-stress}
\end{figure}

To further investigate these relaxation dynamics, we study the inter-particle interaction forces.
In our simulations we can directly sample the instantaneous force contribution of the particle-particle interactions, i.e. stemming from equation~\ref{eq:fluidstructure:interactions} without any of the additional force terms.
We do not expect any interactions through the fluids, since the movement of the particles is very slow, and therefore no significant pressure contributions are expected in between the particles due to compressions in the fluid.
Also, no significant capillary interactions as studied in our previous paper \cite{WASH20} are expected, since the height of the fluid-fluid interface is positioned near the centre of the centre of mass of the particles.

In Fig.~\ref{fig:voronoi-stress} we qualitatively project the total interaction force experienced by each particle onto the Voronoi tessellation of the system in the $xy$-plane sliced at the mean height of the center of mass of the particles.
Here, it can be observed that long chains of high interaction force are formed for the rigid particles.
When the particles become softer, the interaction force between particles is more homogeneously spread, and the long force chains become patches where the increase of the particle interaction force occurs more gradual.

\begin{figure}
    \centering\includegraphics[width = .9\textwidth]{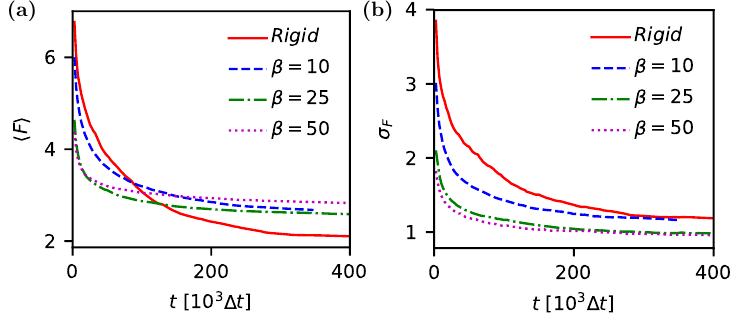}
    \vspace{-3mm}
    \caption[Particle interaction force in a layer of soft particles]{
        Relaxation of (left panel) the mean and (right panel) the standard deviation of the total particle-particle interaction force per particle for different softness parameters. The total number of particles is $N=4445$.
    }
    \label{fig:total-stress}
\end{figure}
Figure~\ref{fig:total-stress} shows both the mean particle-particle interaction force, and the standard deviation of this force.
In the beginning of the simulation the average interaction force is largest for stiffer particles, causing a faster relaxation towards an ordered packing.
In agreement with the observations made so far, the standard deviation of the interaction force indeed increases when the particles become stiffer, where the perfectly rigid particles have the largest standard deviation.

At the end of our simulations the ratio of the standard deviation and the mean of the interaction force is lower for softer particles. This fact confirms the picture that in the rigid particle system narrow chains with a relatively high interaction force span the whole system, while the particles in between these chains have a relatively low interaction force.
Adding softness to the particles allows them to dissipate more efficiently such strong forces, as suggested by their smaller standard deviation.

When comparing Figs.~\ref{fig:voronoi-stress} and \ref{fig:voronoi-equilateral}, it becomes apparent that these force chains are located mainly in between the regions with hexagonally packed particles.
This agrees with the expectation that our system relaxes towards a homogeneous, ordered system, where the presence of defects increases its energy.

\section{Evaporation rate of soft particle suspensions}\label{sec:time}

When particles are suspended homogeneously throughout to the evaporating fluid, over time some will adsorb onto the fluid interface, and the evaporation rate slows down as a result of the reduction of the excess free surface of the fluid-fluid interface.
As long as the total fraction of particles in the fluid is ${\Phi} \leq {\Phi_0}$, the evaporating fluid component can easily flow through gaps between the particles.
However, when the packing fraction increases further above ${\Phi_0}$, these inter-particle gaps become partially blocked by a second layer of particles forming underneath the layer of particles adsorbed onto the fluid-fluid interface.
Consequently, the evaporating fluid component has to flow through this maze-like configuration of the particles.
The actual packing of the particles plays a role on how easily the fluid can flow through the particle layer near the interface, where large defects in the particle packing can be expected to enhance the evaporation rate as compared to a tight, homogeneous particle packing.

As a first study of an evaporating system, we initialise a system of $D=200\times{}200\times{}80$ fluid nodes, with a rigid substrate at the bottom plane, and evaporative boundary conditions at the top plane with ${\rho_{\mathrm{evap}}}=0$.
We initialise a thick fluid film with ${h_0}=8{\hat{R}}$ of the red fluid component at the bottom of the system, and fill the remainder with the blue component.
The fraction ${\Phi}$ of particles in the system is characterised relative to the theoretical packing fraction ${\Phi_0}$ needed to cover the fluid-fluid interface with a single layer of particles.
The evaporation rate is enforced by applying an evaporative boundary condition on the top plane with ${\rho_{\mathrm{evap}}}=0$. The total number of particles corresponding to $\Phi_0$ is $N=192$.

\begin{figure}
    \centering
    \includegraphics[width = .85\textwidth]{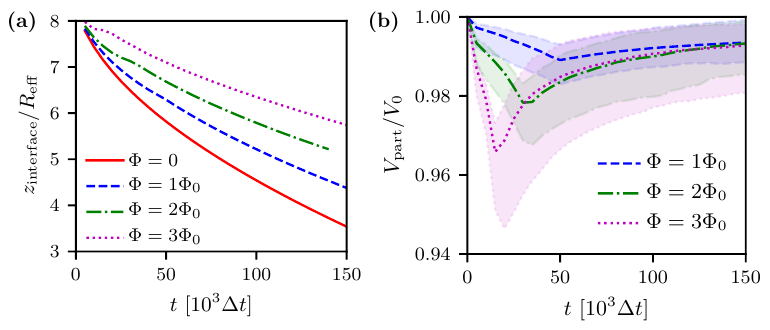}
    \caption[Drying rate of soft particle suspension]{
        Drying of a soft particle suspension with ${\beta}=50$, ${\rho_{\mathrm{evap}}}=0$, and an initial film height of $8{\hat{R}}$ for different packing fractions.
        Left panel: average instantaneous height of the fluid-fluid interface. 
        Right panel: mean instantaneous particle volume (thick lines) for different packing fractions. The shaded regions span the region between the lowest and highest instantaneous particle volume. The total number of particles corresponding to $\Phi_0$ is $N=192$.
    }
    \label{fig:drying-time-particles}
\end{figure}

In figure~\ref{fig:drying-time-particles}a, we show the time evolution of the interface height for different particle packing fractions and compare the results to the evaporation of a fluid without particles. As expected, the evaporation rate reduces for an increasing number of particles present in the system. We observe a small deviation from an otherwise smoothly decreasing interface height, which is larger and occurs earlier for increasing packing fractions. For a particle fraction of ${\Phi}=2{\Phi_0}$ this deviation appears around $t={2.5\times{}10^4}\Delta t$, whereas for ${\Phi}=3{\Phi_0}$ it already appears around $t={1.5\times{}10^4}\Delta t$. This change in the evaporation rate could be related to the deformation of the particles. In fact, in Figure~\ref{fig:drying-time-particles}b one can observe the time evolution of the instantaneous volume of the particles. In the initial state, all particles have the same volume of $V_0$. When a particle adsorbs at the fluid-fluid interface, the surface tension exerts a tensile force on the particle boundary, and thereby reduces its volume. As a result of the elastic properties of its surface, the particle then decompresses trying to expand back towards its initial volume $V_0$.   The minimum in the particle volume occurs around the same time as the deviation observed in the evolution of the interface height, suggesting that the temporarily decreased evaporation rate results from decompressing particles. This effect occurs earlier in the simulation and is stronger for larger packing fractions because more particles are initialised close to the interface.

The influence of this compression and decompression on the position of the interface can be understood as follows:
When a particle centred at the interface decompresses, i.e. its volume increases, the finite system size requires the volumes filled with blue or red fluid to reduce by the same amount. Incompressibility and volume conservation of the fluids cause an increase of the pressure on both sides of the interface which is proportional to the size of the individual fluid domains.  
Hence, to equilibrate the pressure between the regions filled with red and blue component, the interface moves downwards towards the red fluid volume when the particle volume increases. Similarly, the interface moves upwards towards the blue fluid when the particles compress at the interface.

If many particles adsorb onto a fluid-fluid interface simultaneously, the position of the interface raises first, after which it lowers again when the particle volume equilibrates.
For larger packing fractions, more particles are initialised close to the interface, and the rate at which particles adsorb onto the interface is much faster in comparison to a system with a lower packing fraction.
Hence, the effect of the compression and decompression cycle of the particle adsorption is stronger and occurs sooner in our simulations for increasing packing fractions. The fastest evaporation rate in our simulations happens at $4 R  / 5\times10^4 \Delta t$. For water, assuming a surface tension and viscosity of order $1$ (about $100$ \emph{mN/m} and $1$ \emph{mN s / m$^2$}, respectively) and a colloid of radius $R = 100$ \emph{nm} leads to a $\Delta x \approx 10$ \emph{nm} and $\Delta t \approx 10^{-10}$ \emph{s}. The estimated evaporation rate is $v = 4 \times 100$ \emph{nm}  $/ (5\times 10^4\times 10^{-10}$ \emph{s}) $\simeq 0.1$ \emph{m/s}.

\section{Drying of thin films with soft particles}\label{sec:drying}
In Sec.~\ref{sec:relaxation}, we have studied the relaxation and ordering of a single layer of soft particles, which were initialised randomly in a thin fluid film. 
In this section, instead, we report on the dynamics of an evaporating suspension of soft particles using evaporative boundary conditions, which allows us to change the evaporation rate in order to study the influence of the diffusive and evaporative time scales on the drying process.

We initialise our system in the same way as in Sec.~\ref{sec:relaxation}. All simulations are initialised with 4806 particles, corresponding to the close-packing fraction ${\Phi} \simeq {\Phi_0}$ for rigid discs.
The systems are sufficiently dilute such that we do not need a growing stage during the initialisation, and we can simply place the centre of mass position of the particles at positions randomly drawn from a homogeneous distribution. 
We apply the evaporative boundary conditions with ${\rho_{\mathrm{evap}}}=0$ at the top of the simulation domain, while the bottom is still occupied by the solid substrate.

Initially, particles start adsorbing on to fluid-fluid interface, which is lowering because of the evaporation of the lower fluid. During the initial phase of the drying, there is plenty of excess surface area for the particles to adsorb onto, but as a result of the relatively small film height, the fluid-fluid interface quickly becomes jammed with particles.

After roughly $2.5\times{}10^4$ simulation steps, the entire interface is occupied with particles, and plenty of particles are jammed.
During the evaporation, the still freely moving particles below the interface become jammed between the particle-laden interface and the bottom substrate, resulting in the fluid interface changing from a semi-planar structure into an undulating surface.

\begin{figure}
    \centering\includegraphics[width = .75\textwidth]{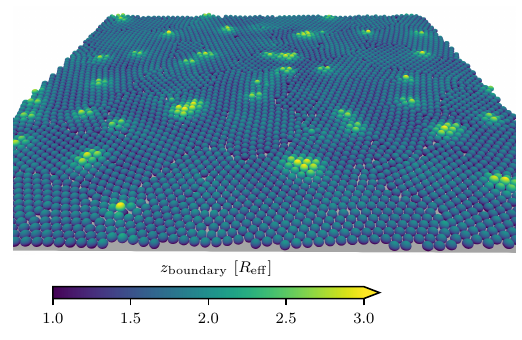}
    \vspace{-3mm}
    \caption[Visualisation of dried suspension with ${\beta}=10$]{
        Snapshot of a simulation of 4806 particles with ${\beta}=10$ after $t=5\times{}10^4$ steps. The boundary elements of the particles are coloured according to their height.
        For clarity, the fluid-fluid interface is not shown. 
    }
    \label{fig:example-drying}
\end{figure}

In figure~\ref{fig:example-drying} we visualise an instantaneous snapshot of a system with $4806$ particles and ${\beta}=10$ at $t=5\times{}10^4\Delta t$.
Here, defects in the particle packing can be observed similarly to what is seen in section~\ref{sec:relaxation}.
However, at multiple places there are particles jammed underneath the layer of particles which are adsorbed onto the fluid-fluid interface.

For an analysis similar to the one in section~\ref{sec:relaxation}, we can here not use all particles in the system, nor simply extract all particles within a given height band, since this then includes both particles adsorbed onto the interface as particles jammed underneath.
Therefore, we use a progressive filtering algorithm, which removes all particles not adsorbed onto the fluid interface.

\begin{figure}
    \centering\includegraphics[width = .9\textwidth]{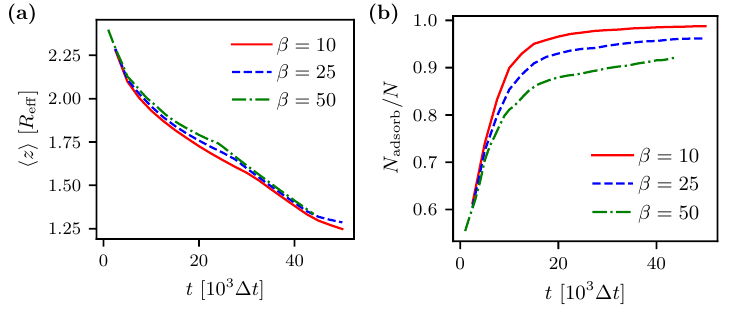}
    \vspace{-3mm}
    \caption[Evolution of number of adsorbed particles and their average height]{
        Time evolution of mean elevation (left panel) and fraction of adsorbed particles at the fluid interface for a drying suspension with $N=4806$ soft particles in a domain of $1000\times{}1000\times{} 32$ lattice units.
        The fluid film has an initial height of $3{R_0}$, and evaporates due to the imposed evaporation boundary conditions with ${\rho_{\mathrm{evap}}}=0$ at the top of the system. 
        Right panel: Number of particles adsorbed onto the fluid-fluid interface.
        No significant change is observed in the rate at which the particle-laden interface lowers, while the number of particles adsorbed onto the interface reduces with the particle softness as a result of the increased stretching of soft particles at the interface.
        Hence, while less particles adsorb onto the fluid-fluid interface, the uncovered surface area remains roughly equal.
    }
    \label{fig:drying-softness-trapped}
\end{figure}

During drying the particles get trapped underneath the fluid interface.
Figure~\ref{fig:drying-softness-trapped}a shows the evolution of the average height of the centre of mass of the particles adsorbed onto the fluid-fluid interface.
The fluid film appears to evaporate slower when the particles adsorbed onto it are softer.
This difference is expected, since an increased softness allows particles to fill the gaps in between particles better, reducing the excess free area of the interface.
However, the observed differences are relatively small, indicating that there is no significant decrease in the excess free area for increasing softness parameters.

This results from the relatively large separation distance between the particles due to the used particle-particle interaction force.
In a 2D plane, an ideal packing of circles has about 9\% of the total area unoccupied by particles.
However, of the remaining area, roughly 18\% is not enclosed by a particle boundary as a result of the relatively large interaction range, and is therefore also not occupied by particles.
Extremely soft particles are capable to deform such as to fill the former free area entirely, but the remaining area can only be reduced by reducing the interaction range.
Hence, it is expected that for the small range of softness parameters used, no significant decrease is observed in the drying rate.

While the average height of the particles adsorbed onto the interface only shows a minor difference for different particle softnesses, in figure~\ref{fig:drying-softness-trapped}b we do observe a significant difference in the number of particles that are adsorbed onto the fluid-fluid interface.
For an increasing softness parameter less particles adsorb onto the interface, and consequently more particles remain trapped underneath the interface.
This results directly from the softness of the particles, where softer particles occupy a larger surface area of the interface, and the equal number of particles in all simulations.
The increasing amount of particles trapped underneath the particle-laden interface also influences the evaporation rate, as was already shown in the previous section.
However, for these systems the amount of particles trapped underneath the interface is relatively low, and we expect no significant influence of the trapped particles on the evaporation rate.

During the drying of the suspension, defects can form in the particle packing, as observed in figure~\ref{fig:example-drying}.
Similarly as performed in section~\ref{sec:relaxation}, we quantify the presence of these defects by using a Voronoi tessellation of the particle positions, and characterise the ordering by the Voronoi entropy and isotropic order parameter.

\begin{figure}
    \centering\includegraphics[width = .75\textwidth]{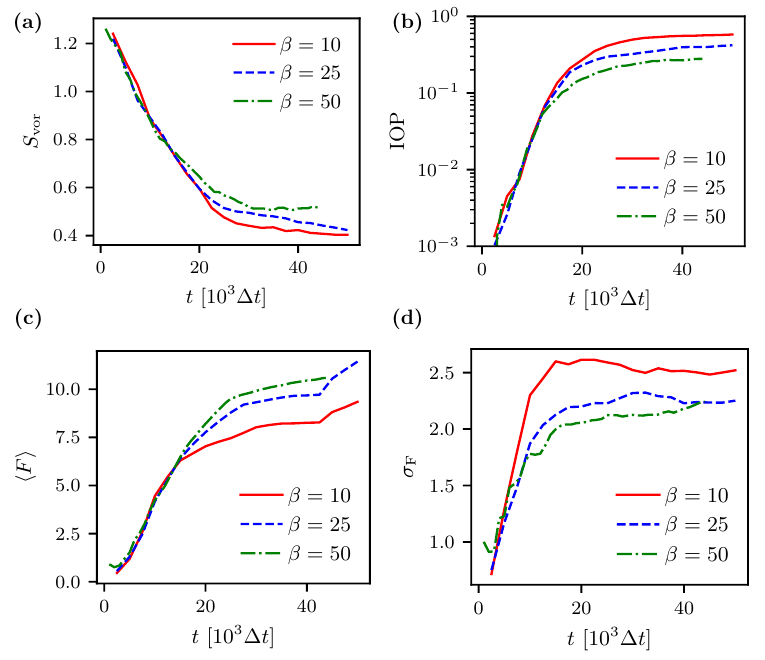}
    \vspace{-3mm}
    \caption[Order parameters for drying suspensions with different softness]{
        Order parameters and inter-particle interaction force for a drying suspension with $N=4806$ soft particles in a domain of $1000\times{}1000\times{}32$ lattice units. 
        The fluid film has an initial height of 3\ensuremath{R_0}, and evaporates due to the imposed boundary conditions with \ensuremath{\rho_{\mathrm{evap}}}=0.0 at the top of the system. From top to bottom, left to right:
        (a) Voronoi entropy for different softness parameters.
        (b) Isotropic order parameter for different softness parameters.
        (c) Mean and (d) standard deviation of the particle-particle interaction force experienced by particles adsorbed onto the interface.
    }
    \label{fig:drying-ordering}
\end{figure}

Figure~\ref{fig:drying-ordering}a shows the evolution of the Voronoi entropy of the particles adsorbed onto the fluid interface.
In the first stage, up to $t \simeq 2\times{}10^4 \Delta t$, the Voronoi entropy does not show any significant difference for the different softness parameters.
In this stage, the fluid interface is not yet saturated with particles, and as a result of the same initial state for the simulations with different softness parameters the rate at which particles adsorb onto the fluid-fluid interface is approximately equal.

After $t \simeq 2\times{}10^4 \Delta t$ the fluid-fluid interface is almost completely packed with particles, as was seen in figure~\ref{fig:drying-softness-trapped}b where the number of adsorbed particles saturates.
From this point in time differences appear in the Voronoi entropy for different softness parameters.
Where for the first stage the adsorption of new particles onto the interface drove the ordering of the system, now the particles re-order as a result of their elastic behaviour and the interaction forces between particles.
Similar as was observed in section~\ref{sec:relaxation}, the stiffer particles again reach a more ordered particle packing with a lower Voronoi entropy than the softer particles.

A similar observation can be made for the isotropic order parameter, which is shown in figure~\ref{fig:drying-ordering}b.
First, the IOP grows rapidly, with no differentiation between the softness parameters.
Afterwards it saturates, where the stiffer particles show an overall more ordered particle packing as compared to softer particles.
In figure~\ref{fig:drying-ordering}c-d we characterise the interaction force between the particles adsorbed onto the fluid-fluid interface.
When the particle-laden interface does not yet interact with the substrate through the trapped particles, no significant difference is observed for the different softness parameters.
However, afterwards softer particles show a larger average interaction force, and a smaller standard deviation of the interaction force, than stiffer particles.
This can be attributed to the increase in the number of particles trapped underneath the interface.
Similarly, the standard deviation of the interaction force also only starts to differentiate for ${\beta}$ in the second stage.
Here, softer particles tend to have a more homogeneously spread interaction force as compared to stiffer particles, similar to section~\ref{sec:relaxation} for a single layer of soft particles.

The rapid increase of the mean interaction force after roughly $4\times{}10^4$ simulation steps directly results from the interaction between particles and the substrate.
As seen from figure~\ref{fig:drying-softness-trapped}a, the height of the particle film becomes comparable to the radius of the particle.
Hence, the interaction between the particle and the substrate starts to play a significant role, and causes the sudden rise in the average interaction force.
This additional interaction occurs earlier and becomes more significant for softer particles, since more particles become trapped underneath the particle-laden fluid interface.
Since the interface is relatively flat, and most particles are adsorbed onto it, the standard deviation of the particle interaction force is not significantly impacted by the increasing magnitude of the particle-substrate interaction force.

\section{Conclusion}\label{sec:conclusion}
We reported on drying soft particle suspensions, simulated using the lattice Boltzmann method, coupled to a finite-element description of the particle interface. First, we studied how particle softness influences the final stage of drying, by investigating the relaxation of particles initially arranged in a monolayer on top of a substrate. Here, the main driving force for the ordering of the particles is the particle-particle interaction force. Perfectly rigid particles form long, narrow chains through which the interparticle force and, consequently, the stress, is distributed. For increasing particle softness, these correlated regions of intense stress become broader and less pronounced, albeit always located at the defects. An ordered, hexagonal packing typically characterizes the regions between these chain-like structures. 
Next, we studied the drying rate of a soft particle suspension using evaporation boundary conditions. Here, the increase of the packing fraction reduces the evaporation rate of the fluid film significantly, because the solvent needs to flow through the narrow interstices between the tightly packed particles. Finally, we studied the influence of the particle softness on the drying rate. One could expect that softer particles would contribute to a decrease in the drying rate since they are better capable of filling the gaps where multiple particles meet at the interface. However, the free surface area in between two particles turns out to make up a significant portion of the total free surface area of the fluid-fluid interface. Hence, only a marginal influence was observed for the softness parameter on the drying rate.

\begin{acknowledgments}
{We acknowledge financial support from the Dutch Research Council NWO/TTW (project 10018605), HPC-Europa3 (Grant INFRAIA-2016-1-730897), and the German Research Foundation DFG (research unit FOR2688, project HA4382/8-1). The authors thank the Jülich Supercomputing Centre and the High Performance Computing Centre Stuttgart for the allocated CPU time.}
\end{acknowledgments}

\enlargethispage{20pt}


\begin{thebibliography}{37}
\makeatletter
\providecommand \@ifxundefined [1]{%
 \@ifx{#1\undefined}
}%
\providecommand \@ifnum [1]{%
 \ifnum #1\expandafter \@firstoftwo
 \else \expandafter \@secondoftwo
 \fi
}%
\providecommand \@ifx [1]{%
 \ifx #1\expandafter \@firstoftwo
 \else \expandafter \@secondoftwo
 \fi
}%
\providecommand \natexlab [1]{#1}%
\providecommand \enquote  [1]{``#1''}%
\providecommand \bibnamefont  [1]{#1}%
\providecommand \bibfnamefont [1]{#1}%
\providecommand \citenamefont [1]{#1}%
\providecommand \href@noop [0]{\@secondoftwo}%
\providecommand \href [0]{\begingroup \@sanitize@url \@href}%
\providecommand \@href[1]{\@@startlink{#1}\@@href}%
\providecommand \@@href[1]{\endgroup#1\@@endlink}%
\providecommand \@sanitize@url [0]{\catcode `\\12\catcode `\$12\catcode
  `\&12\catcode `\#12\catcode `\^12\catcode `\_12\catcode `\%12\relax}%
\providecommand \@@startlink[1]{}%
\providecommand \@@endlink[0]{}%
\providecommand \url  [0]{\begingroup\@sanitize@url \@url }%
\providecommand \@url [1]{\endgroup\@href {#1}{\urlprefix }}%
\providecommand \urlprefix  [0]{URL }%
\providecommand \Eprint [0]{\href }%
\providecommand \doibase [0]{http://dx.doi.org/}%
\providecommand \selectlanguage [0]{\@gobble}%
\providecommand \bibinfo  [0]{\@secondoftwo}%
\providecommand \bibfield  [0]{\@secondoftwo}%
\providecommand \translation [1]{[#1]}%
\providecommand \BibitemOpen [0]{}%
\providecommand \bibitemStop [0]{}%
\providecommand \bibitemNoStop [0]{.\EOS\space}%
\providecommand \EOS [0]{\spacefactor3000\relax}%
\providecommand \BibitemShut  [1]{\csname bibitem#1\endcsname}%
\let\auto@bib@innerbib\@empty


\bibitem{Keddie2010}
Keddie, J. \& Routh, A., 2010 \emph{Fundamentals of latex film formation:
  processes and properties}.
\newblock Springer

\bibitem{Kooij2015}
van~der Kooij, H. \& Sprakel, J., 2015 Watching paint dry; more exciting than
  it seems, \emph{Soft Matter} \textbf{11}, 6353--6359

\bibitem{Dufresne2003}
Dufresne, E.~R., Corwin, E.~I., Greenblatt, N.~A., Ashmore, J., Wang, D.~Y.,
  Dinsmore, A.~D., Cheng, J.~X., Xie, X.~S., Hutchinson, J.~W. \& Weitz, D.~A.,
  2003 Flow and fracture in drying nanoparticle suspensions, \emph{Phys. Rev.
  Lett.} \textbf{91}, 224501

\bibitem{Singh2007}
Singh, K. \& Tirumkudulu, M., 2007 Cracking in drying colloidal films,
  \emph{Phys. Rev. Lett.} \textbf{98}, 218302

\bibitem{Roberts2012}
Roberts, C.~C. \& Francis, L.~F., 2013 Drying and cracking of soft latex
  coatings, \emph{J. of Coatings Tech. and Res.} \textbf{10}, 441--451

\bibitem{Goehring2013}
Goehring, L., Clegg, W. \& Routh, A., 2013 Plasticity and fracture in drying
  colloidal films, \emph{Phys. Rev. Lett.} \textbf{110}, 024301

\bibitem{Man2008}
Man, W. \& Russel, W., 2008 Direct measurements of critical stresses and
  cracking in thin films of colloid dispersions, \emph{Phys. Rev. Lett.}
  \textbf{100}, 198302

\bibitem{Piroird2016}
Piroird, K., Lazarus, V., Gauthier, G., Lesaine, A., Bonamy, D. \& Rountree,
  C.~L., 2016 Role of evaporation rate on the particle organization and crack
  patterns obtained by drying a colloidal layer, \emph{Europhys. Lett.}
  \textbf{113}, 38002

\bibitem{Winnik1997}
Winnik, M., 1997 Latex film formation, \emph{Curr. Opin. Colloid Interface
  Sci.} \textbf{2}, 192--199

\bibitem{Routh2001}
Routh, A.~F. \& Russel, W.~B., 2001 Deformation mechanisms during latex film
  formation: Experimental evidence, \emph{Industrial \& engineering chemistry
  research} \textbf{40}, 4302--4308

\bibitem{MHS16}
Mehrabian, H., Harting, J. \& Snoeijer, J.~H., 2016 Soft particles at a fluid
  interface, \emph{Soft Matter} \textbf{12}, 1062--1073

\bibitem{Trueman2012}
Trueman, R., Lago~Domingues, E., Emmett, S., Murray, M. \& Routh, A., 2012
  Auto-stratification in drying colloidal dispersions: A diffusive model,
  \emph{J. Colloid Interf. Sci.} \textbf{377}, 207--212

\bibitem{Fujita2007}
Fujita, M. \& Yamaguchi, Y., 2007 Multiscale simulation method for
  self-organization of nanoparticles in dense suspension, \emph{J. Comput.
  Phys.} \textbf{223}, 108--120

\bibitem{Fujita2015}
Fujita, M., Koike, O. \& Yamaguchi, Y., 2015 Direct simulation of drying
  colloidal suspension on substrate using immersed free surface model, \emph{J.
  Comput. Phys.} \textbf{281}, 421--448

\bibitem{Uzi2015}
Uzi, A., Ostrovski, Y. \& Levy, A., 2015 Modeling and simulation of mono-layer
  coating, \emph{Drying Technology} \textbf{33}, 1798--1807

\bibitem{Zhao2019}
Zhao, M., Luo, W. \& Yong, X., 2019 Harnessing complex fluid interfaces to
  control colloidal assembly and deposition, \emph{J. Colloid Interf. Sci.}
  \textbf{540}, 602--611

\bibitem{Liang2013}
Liang, G., Zeng, Z., Chen, Y., Onishi, J., Ohashi, H. \& Chen, S., 2013
  Simulation of self-assemblies of colloidal particles on the substrate using a
  lattice {Boltzmann} pseudo-solid model, \emph{J. Comput. Phys.} \textbf{248},
  323--338

\bibitem{Wouters19}
Wouters, M. P.~J., Aouane, O., Kr\"uger, T. \& Harting, J., 2019 Mesoscale
  simulation of soft particles with tunable contact angle in multi-component
  fluids, \emph{Phys. Rev. E} \textbf{100}, 033309

\bibitem{WASH20}
Wouters, M. P.~J., Aouane, O., Sega, M. \& Harting, J., 2020 Capillary
  interactions between soft particles protruding through thin fluid films,
  \emph{Soft matter} \textbf{16}, 10910--10920

\bibitem{benzi1992lattice}
Benzi, R., Succi, S. \& Vergassola, M., 1992 The lattice {Boltzmann} equation:
  theory and applications, \emph{Phys. Rep.} \textbf{222}, 145--197

\bibitem{kruger2017lattice}
Kr{\"u}ger, T., Kusumaatmaja, H., Kuzmin, A., Shardt, O., Silva, G. \& Viggen,
  E.~M., 2017 \emph{The Lattice {B}oltzmann Method}.
\newblock Springer

\bibitem{bhathnagor1954model}
Bhathnagar, P., Gross, E. \& Krook, M., 1954 A model for collision processes in
  gases, \emph{Phys. Rev.} \textbf{94}, 511

\bibitem{qian1992lattice}
Qian, Y., d'Humi{\`e}res, D. \& Lallemand, P., 1992 Lattice {BGK} models for
  {N}avier-{S}tokes equation, \emph{Europhys. Lett.} \textbf{17}, 479

\bibitem{shan1993lattice}
Shan, X. \& Chen, H., 1993 Lattice {B}oltzmann model for simulating flows with
  multiple phases and components, \emph{Phys. Rev. E} \textbf{47}, 1815

\bibitem{hessling2017diffusion}
Hessling, D., Xie, Q. \& Harting, J., 2017 Diffusion dominated evaporation in
  multicomponent lattice {B}oltzmann simulations, \emph{J. Chem. Phys.}
  \textbf{146}, 054111

\bibitem{skalak1973strain}
Skalak, R., Tozeren, A., Zarda, R. \& Chien, S., 1973 Strain energy function of
  red blood cell membranes, \emph{Biophys. J.} \textbf{13}, 245

\bibitem{kruger2011efficient}
Kr{\"u}ger, T., Varnik, F. \& Raabe, D., 2011 Efficient and accurate
  simulations of deformable particles immersed in a fluid using a combined
  immersed boundary lattice {B}oltzmann finite element method, \emph{Comp.
  Math. App.} \textbf{61}, 3485--3505

\bibitem{meyer2003discrete}
Meyer, M., Desbrun, M., Schr{\"o}der, P. \& Barr, A.~H., 2003 Discrete
  differential-geometry operators for triangulated 2-manifolds.
\newblock \emph{Visualization and mathematics III}, pages 35--57. Springer

\bibitem{biben2011three}
Biben, T., Farutin, A. \& Misbah, C., 2011 Three-dimensional vesicles under
  shear flow: Numerical study of dynamics and phase diagram, \emph{Phys. Rev.
  E} \textbf{83}, 031921

\bibitem{barthes2016motion}
Barth\`{e}s-Biesel, D., 2016 \emph{Motion and deformation of elastic capsules and vesicles in flow}.
\emph{Annu. Rev. Fluid Mech.} \textbf{48}, 25--52

\bibitem{wouters2020mesoscopic}
Wouters, M., 2020 \emph{Mesoscopic modelling of drying phenomena of soft particle suspensions}.
\newblock PhD thesis, TU Eindhoven

\bibitem{KKH14}
Krüger, T., Kaoui, B. \& Harting, J., 2014 Interplay of inertia and
  deformability on rheological properties of a suspension of capsules, \emph{J.
  Fluid Mech.} \textbf{751}, 725--745

\bibitem{HKH11}
Hyv\"aluoma, J., Kunert, C. \& Harting, J., 2011 Simulations of slip flow on
  nanobubble-laden surfaces, \emph{J. Phys.: Cond. Matt.} \textbf{23}, 184106

\bibitem{HVC04}
Harting, J., Venturoli, M. \& Coveney, P.~V., 2004 Large-scale grid-enabled
  lattice-Boltzmann simulations of complex fluid flow in porous media and under
  shear, \emph{Phil. Trans. R. Soc. London Series A} \textbf{362}, 1703--1722

\bibitem{SSFKSHC17}
Schmieschek, S., Shamardin, L., Frijters, S., Krüger, T., Schiller, U.~D.,
  Harting, J. \& Coveney, P.~V., 2017 Lb3d: A parallel implementation of the
  lattice-Boltzmann method for simulation of interacting amphiphilic fluids,
  \emph{Comp. Phys. Comm.} \textbf{217}, 149--161

\bibitem{ASH20}
Aouane, O., Scagliarini, A. \& Harting, J., 2021 Structure and rheology of
  suspensions of spherical strain-hardening capsules, \emph{J.
  Fluid Mech.} \textbf{911}, A11

\bibitem{Limaye1996}
Limaye, A.~V., Narhe, R.~D., Dhote, A.~M. \& Ogale, S.~B., 1996 Evidence for
  convective effects in breath figure formation on volatile fluid surfaces,
  \emph{Phys. Rev. Lett.} \textbf{76}, 3762--3765

\end{thebibliography}
\end{document}